# Designing Model and Optimization of the Permanent Magnet for Joule Balance NIM-2

Q. You, J. Xu, Z. Li and S. Li

*Abstract* — Permanent magnets with yokes are widely used in the watt and joule balances to measure the Planck constant for the forthcoming redefinition of the unit of mass, the kilogram. Recently, a permanent magnet system has been in consideration for a further practice of NIM-2, the generalized joule balance. In this paper, an analytical model to design the permanent magnet system is presented. The presented model can be solved to obtain the preliminary parameters and then is used as guidance for FEA software to optimize the parameters of such magnetic system. As a instance for the application of the designing model, the design of the permanent magnet system for NIM-2 is described and the special design of opening shape makes the misalign of the top and middle yokes has little influence on the vertical component of the magnetic field.

## I. INTRODUCTION

Electromechanical balances, e.g. watt balance [1] and joule balance [2], are currently used in majority of the national metrology institutes (NMIs) for precisely measuring the Planck constant ($h$), which is one of the most promising methods for the redefinition of the kilogram [3-4].

A stable and strong magnetic field is a basic element in both watt and joule balances. Typically, a permanent magnet system is desired and the magnetic circuit design is usually with the help of the finite element analysis (FEA) software [5-7]. However, FEA method ignores the physical principles behind the design, showing limited design and optimization ideas. Indeed, an analytical model will show the design idea clearly, which can be used as the guidance before the detailed simulations are calculated by FEA software.

The basic expressions of an analytical permanent-magnet design model have been described to summarize the physical principles of the permanent magnet system in watt and joule in [8]. In this paper, the expressions of this model is derived, including the optimal ($BL$), the magnetic flux density in the air gap, dimension of the yokes and the force to open the yokes. Besides, how to adjust the uniformity of the magnetic field is also discussed.

Magnetic flux linkage method is used for Joule balance NIM-2. At present, an electromagnet system is used to produce the needed magnetic field, which could work well at $10^{-7}$ level measurement. For an alternative consideration, a permanent magnet is also in consideration to decrease the effect from the nonlinearity and heating problem of the electromagnet. As an instance of the analytical designing model, the permanent magnetic system for NIM-2 is designed and described in section III.

## II. PERMANENT MAGNET DESIGNING MODEL

### A. The Optimal Value of (BL)

A decision of the geometrical factor ($BL$) needs to be made, an integral of the magnetic flux density along the coil wire, to suppress the relative combined uncertainty in the two measurement modes of both watt and joule balances. For joule balance method, the principle formula is

$$mgz + \Delta W = \frac{V}{R}\Delta \varphi, \qquad (1)$$

where $R$ is the sampling resistance of the coil current and $V$ is the voltage of the sampling resistance. Then the mass ($m$) can be calculated as

$$m = \frac{\frac{V}{R}\Delta\varphi - \Delta W}{gz} = \frac{K}{gz}, \qquad (2)$$

where

$$K = \frac{V}{R}\Delta\varphi - \Delta W = mgz. \qquad (3)$$

Based on the uncertainty propagation formula, we obtain

$$\frac{\sigma_m^2}{m^2} = \frac{\sigma_K^2}{K^2} + \frac{\sigma_g^2}{g^2} + \frac{\sigma_z^2}{z^2} \qquad (4)$$

where $\sigma_K$ is the combined uncertainty of $K$ and is also calculated by the uncertainty propagation formula as

Manuscript received July. 12, 2016. This work is supported in part by the key support program of Major research project of China National Natural Science Foundation (Grant No.91536224) and in part by the China National Key Scientific Instrument and Equipment Development Project (Grant No.2011YQ090004)

Q. You is with Tsinghua University (phone: +86-10-64526190; e-mail: neroyouqiang@163.com).
S. Li, J. Xu are with Tsinghua University, Beijing 100084, China.
Z. Li are with the National Institute of Metrology (NIM), Beijing 100029, China.

$$\frac{\sigma_K^2}{K^2} = \frac{\sigma_K^2}{(mgz)^2} = \frac{(\frac{\sigma_V^2}{V^2} + \frac{\sigma_R^2}{R^2} + \frac{\sigma_{\Delta\varphi}^2}{\Delta\varphi^2})(\frac{V}{R}\Delta\varphi)^2 + \sigma_{\Delta W}^2}{(mgz)^2} \quad (5)$$
$$= \frac{\sigma_V^2}{V^2} + \frac{\sigma_R^2}{R^2} + \frac{\sigma_{\Delta\varphi}^2}{\Delta\varphi^2} + \frac{\sigma_{\Delta W}^2}{(mgz)^2}$$

Ignoring the residual force ($\Delta f$) and assuming that the magnetic field is absolutely uniform, we obtain

$$mg = Bl\frac{V}{R}, \quad (6)$$
$$\Delta\varphi = Blz. \quad (7)$$

Subscribing (5)-(7) into (4) and the uncertainty of the measured mass is obtained as

$$\frac{\sigma_m^2}{m^2} = \frac{\sigma_V^2(Bl)^2}{m^2g^2R^2} + \frac{\sigma_{\Delta\varphi}^2}{(Bl)^2z^2} + \frac{\sigma_R^2}{R^2} + \frac{\sigma_{\Delta W}^2}{(mgz)^2} + \frac{\sigma_g^2}{g^2} + \frac{\sigma_z^2}{z^2}. \quad (8)$$

The geometrical factor ($BL$) appears in the numerator of the first and the denominator of the second term and then the optimal ($BL$) is

$$(Bl)_{op} = \frac{\sigma_{\Delta\varphi}mgR}{z\sigma_V}. \quad (9)$$

The optimal ($BL$) for watt balance method can be calculated by the same way. The derivation can be found in [6] and is expressed as

$$(Bl)_{op} = \frac{\sigma_U mgR}{v\sigma_V}. \quad (10)$$

In practice, the optimal ($Bl$) can be calculated by three parameters: the magnetic flux density in air gap ($B_a$), the radius of the air gap ($r_0$) and turns of the suspended coil ($N$). These parameters have the same linear relationship with ($Bl$) but cause different additive effects. Increasing $r_0$ or $N$ will increase the volume of the coil and increasing $N$ will also produce more additional magnetomotive force in the weighing mode. Besides, the heating power of the coil linearly decreases with $N$ or $r_0$ while quadratically with $B_a$. Therefore, if the current ($Bl$) is less than the optimal ($Bl$), $B_a$ should be increased first, then $r_0$ and $N$ the last.

*B. The Magnetic Flux Density*

Typically, the magnet construction in watt and joule balances is shown in figure 1. The ferromagnetic material should work at the extreme point of $\mu_r - H$ curve, where the magnetic flux density in the yokes is $B_m$, to reduce the influence of nonlinearity and hysteresis [9-10].

The Ampere's circulation theorem is expressed as

$$B_m A_m \cdot R_{ma} = -H_m L_m, \quad (11)$$

where $A_m$ is the area of the permanent magnet, $L_m$ the thickness of the magnet, $H_m$ the magnetic field strength in the permanent magnet and $R_{ma}$ the total reluctance out of the permanent magnet. The recoil curve is expressed as

$$\frac{B_m}{\mu_r} = H_m - H_c, \quad (12)$$

where $H_c$ is the coercive force of the permanent magnet and $\mu_r$ is the recoil permeability. Combining (11) and (12), we obtain

$$B_m A_m \left(\frac{R_{ma}}{L_m} + \frac{1}{\mu_r A_m}\right) = -H_c. \quad (13)$$

Ignoring the reluctance of the yokes, $R_{ma}$ can be calculated as

$$R_{ma} = \frac{L_a}{\mu_0 A_a}. \quad (14)$$

where $L_a$ is the width of the air gap. Besides, the flux conservation is expressed as

$$B_m A_m = B_a A_a, \quad (15)$$

where $B_a$ is the magnetic flux density in the air gap and $A_a$ the developed area of the air gap. Subscribing (14) and (15) into (13), $B_a$ is obtained as

$$B_a \left(\frac{L_a}{L_m \mu_0} + \frac{A_a}{A_m \mu_r}\right) = -H_c. \quad (16)$$

In (15), since $H_c$ can be measured and $\mu_r \approx \mu_0$, $B_a$ should be adjusted by two ratios, $L_a/L_m$ and $A_a/A_m$. Note that $B_a$ is not proportional to $L_m$ or $A_m$. If only $L_m$ or $A_m$ is increased to strengthen $B_a$, the strengthening speed will decay.

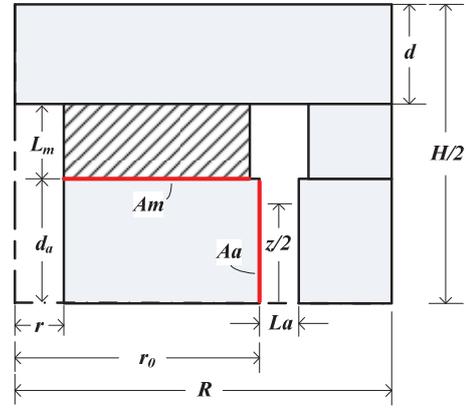

Fig. 1. A quarter section of a typical permanent magnet system in watt and joule balances.

*C. Dimensions of the Yokes*

The horizontal sectional view of the typical permanent magnet is shown in figure 2. For the perfect case, where the ferromagnetic material works at the point with the maximum permeability, the outer yoke will be saturated if it is too narrow and the inner yoke will be saturated if the outer yoke is too wide. Thus the sectional area of the inner yoke should be equal to that of the outer yoke, which is expressed as

$$A_m = (r_o^2 - r^2)\pi = [R^2 - (r_0 + L_a)^2]\pi. \quad (17)$$

where the variations are labeled in Fig. 2.

Similarly, the equivalent area of magnetic path in the top yoke should be equal to the sectional area of the inner or outer yoke, which is expressed as

$$2\pi d \cdot r_{ave} = (r_0^2 - r^2)\pi. \quad (18)$$

where $h$ is the thickness of the top yoke and $r_{ave}$ is the average radius of the magnetic path in the top yoke which can be calculated as $(r_0 + L_a/2)$.

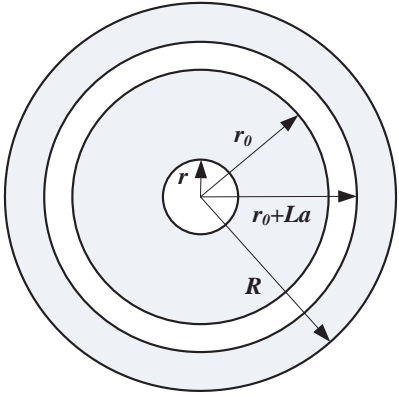

Fig. 2. The horizontal sectional view of the magnetic yokes.

*D. Force to Open the Magnet*

The force to open the magnet can be calculated by the virtual work principle. The calculating model is shown in figure 3. The red dashed line indicates the position to open the magnet. The flux is separated into two parts: one is as flux A shows, which is above the opening position and remains unchanged at the moment when the magnet is opened; another is as flux B shows, which is under the opening position and is cut off when the magnet is opened. The opening force is irrelevant with flux A and thus flux B is named as the effective flux. The magnetic energy of the effective flux covered area is calculated as

$$W_{xm} = \frac{1}{2}\Phi_x^2 R_{xm}. \quad (19)$$

where $R_{xm}$ is the reluctance of the whole magnetic circuit and $\Phi_x$ is the effective flux. $\Phi_x$ can be calculated as

$$\Phi_x = \frac{-H_c L_m}{R_{xm}}. \quad (20)$$

Subscribing (20) into (19) and taking the partial of the energy of $W_x$ with respect to the opening distance $z$, we obtain

$$F_{open} = -\frac{dW_{xm}}{dz} + G = \frac{(H_c L_m)^2}{2R_{xm}^2}\frac{dR_{xm}}{dz} + G. \quad (21)$$

where $G$ is the gravity of the top yoke. At the moment when the magnet is opened, a reluctance of air gap is concatenated into the magnetic circuit and the concatenated reluctance is

$$dR_m = \frac{\mu_0 dz}{S_{x-in}} + \frac{\mu_0 dz}{S_{x-out}} = 2\frac{\mu_0 dz}{S_x}. \quad (22)$$

where $S_{x-in}$ and $S_{x-out}$ is the opening area at the inner and outer yoke, respectively. The $S_{x-in}$ and $S_{x-out}$ is designed to be equal as demonstrated in section C, thus $S_{x-in}$ and $S_{x-out}$ is replaced by $S_x$. Subscribing (22) into (21), we obtain

$$F_{open} = \frac{2\mu_0(H_c L_m)^2}{R_{xm}^2 S_k} + G = \frac{2\mu_0 \Phi_x^2}{S_k} + G. \quad (23)$$

As (22) shows, the opening force is mainly effected by three factors: the effective flux, the opening area and the gravity of the top yoke. When the opening position is decided, increasing the opening area can decrease the opening force. In other word, the "dumpy" magnet has less opening force than the "lanky" magnet. Besides, the opening position can be designed to be closed to the center of the magnet to decrease the opening force on the premise that the uniformity of the magnetic field in the air gap is unaffected.

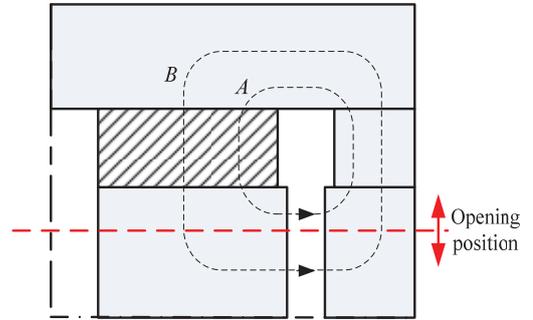

Fig. 3. Calculating model of the force to open the magnet

*E. Adjusting the Uniformity of Magnetic Field*

Adjusting the uniformity of magnetic field in the air gap includes adjusting the horizontal component and adjusting the vertical component of the magnetic field. The horizontal component should be adjusted to be equal while the vertical component should be suppressed. The method of suppressing the vertical component is based on a compensation idea. We can decrease the height of the outer side of the air gap and create some reverse vertical magnetic flux. If the height of the outer side is properly adjusted, the vertical magnetic flux will be suppressed below $1 \times 10^{-3}$T.

The uniformity of the horizontal magnetic field is mainly determined by two factors: the fringe effect and the different length of the magnetic circuits. Because of the fringe effect, the horizontal magnetic field decreases when close to the fringe of the air gap, as the left schematic in figure 4 shows. Inversely, as the right schematic in figure 4 shows, the flux close to the center of the air gap has longer path and thus the horizontal decreases when close to the center of the air gap because of the different length of the magnetic circuit. When the required height of the air gap is decided, the two factors should be adjusted to offset each other to the greatest extent, making the most uniform horizontal magnetic field in the required air gap. This is the reason why the horizontal magnetic field of a well-designed permanent magnet system is always a double peak shaped curve in required air gap.

There are three parameters are often adjusted to improving the uniformity of the horizontal magnetic field: the height of the air gap, the width of the air gap and the permeability of the yokes. Increasing the height of air gap can decreases the fringe effect while the magnetic flux density in air gap will decay based on (16). Narrowing the air gap also decreases the fringe effect but increases the magnetic flux density. It also decreases the ratio of air gap in the magnetic circuit and thus makes the difference of the magnetic circuits bigger. The permeability of yokes directly affects the difference of the magnetic circuits. However, the permeability of yokes is limited by the optional material and thus is not as flexible as directly adjusting the mechanical parameters of the air gap.

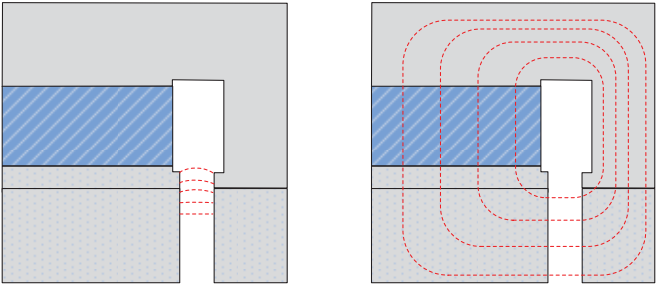

Fig. 4. Calculating model of the force to open the magnet

### F. Solution of the Preliminary Design Parameters

Based on the designing model, the magnet system can be quickly designed for any mechanical sizes. We can obtain the preliminary design parameters from the design model first. An equation set is obtained by combining (9), (16), (17), (18) and (23), which can be solved to obtain the preliminary design parameters of the permanent system. To solve the equation set, other constraints should be added, including the total height of the magnet:

$$H = 2(d + L_m + d_a), \quad (24)$$

and the developed area of the air gap:

$$A_a = 2\pi r_a d_a. \quad (25)$$

Typically, the total height ($H$), the outer radius ($R$) and inner radius ($r$) of the magnet system are mechanically fixed before the design. Besides, the ferromagnetic material of yokes is also known and thus $B_m$ is fixed to make the ferromagnetic material work at with the maximum permeability point. And then there is only one freedom in the equation set. The designer can add one more constraint, such as the opening force or the width of the air gap, to have the equation set be solved.

By solving the equation set, we can get the preliminary design parameters of the magnet system. Then the FEA software is used with the guidance of the designing model for further adjusting every parameter and detailed optimization, including the uniformity of magnetic field in the air gap.

## III. DESIGN OF THE PERMANENT MAGNET SYSTEM FOR JOULE BALANCE NIM-2

### A. General Design

The design of the permanent magnet system for NIM-2 is shown in figure 5. The yokes are composed of DT4C (a kind of electro iron). This high-permeability material can improve the uniformity of the magnetic flux density in air gap caused by the different length of the magnet circuit.

The calculated horizontal magnetic flux density in the required air gap is shown in figure 6, which is adjusted according to the method discussed in section II E. As figure 6 shows, this curve is a double peak shape and the relative uniformity is $3.9\times10^{-3}$ in the 120mm height air gap. The three minima, where z=-60mm, 0mm and 60mm, is almost equal and in this case, the two main factors affecting the uniformity (the fringe effect and the different length of magnetic circuits) can offset each other to the most extent. Besides, the outer side of the air gap is 10mm shorter than the inner side and thus the vertical component of the magnetic flux density is adjusted below $3.9\times10^{-4}$T.

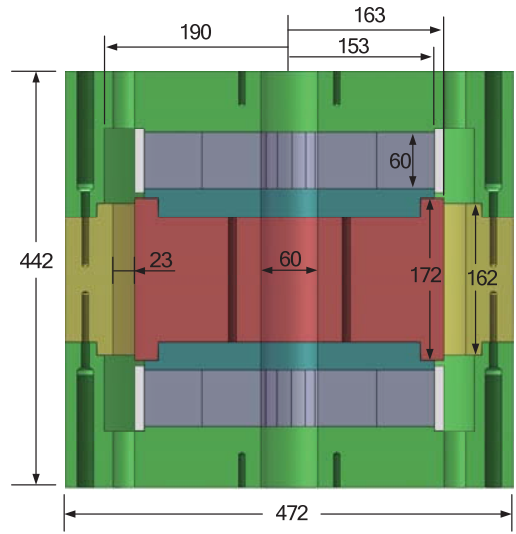

Fig. 5. Cross-sectional view of the permanent magnet system for NIM-2

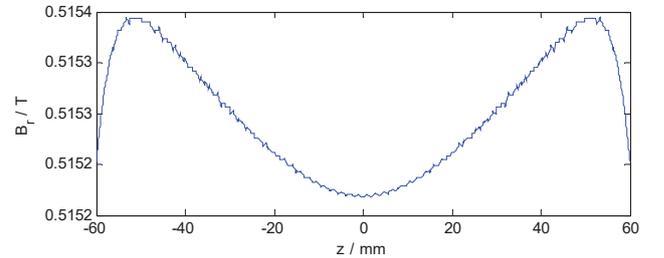

Fig. 6. Calculated horizontal magnetic flux density in the required air gap of the permanent magnet system for NIM-2. The abscissa is the distance from the center of air gap.

### B. Design of the Opening shape

The common design for the opening shape of the permanent magnet system is shown in figure 7 (a). This kind of shape is easy for processing but has very high requirement for the concentricity of top yoke and middle yoke. Little misalign between the top and middle yoke will generate a strong vertical component ($B_z$) in the required air gap. As shown in figure 8, the peak to peak value of $B_z$ is $2\times10^{-3}$T when the misalign is 0.1mm, which is 10 times bigger than our design. In practice, permanent magnet system needs frequently opening. It is a severe challenge to accurately assemble the magnet system after every opening operation. To overcome this, the permanent magnet system for NIM-2 applies the design in figure 7 (b). This kind of opening shape will increase the difficulty of processing and has higher opening force. But the slit is not across the air gap and the air gap remains intact when the top yoke is taken off. Thus, the misalign of the top and middle yokes has little influence on the required air gap. As shown in figure 9, when the misalign is 0.1mm, the peak to peak value of $B_z$ is $4\times10^{-4}$, which still meets our requirement. Keeping the misalign below 0.1mm is easy for the processing. Besides, the raised edges on middle yoke can be used for position and the center positioning pillar can be replaced by the temperature compensation part.

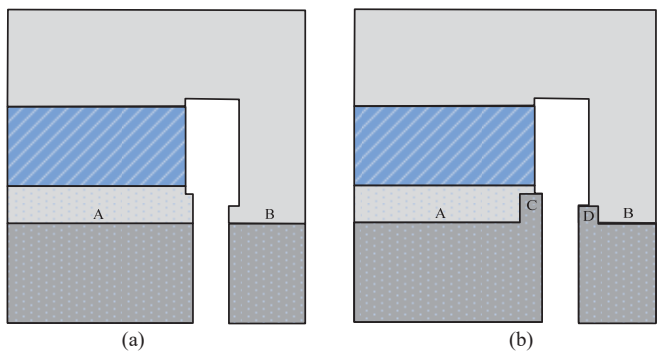

Fig. 7. Comparison of the two designs for opening position. Figure (a) is the typical design and figure (b) is the design in permanent magnet system for NIM-2.

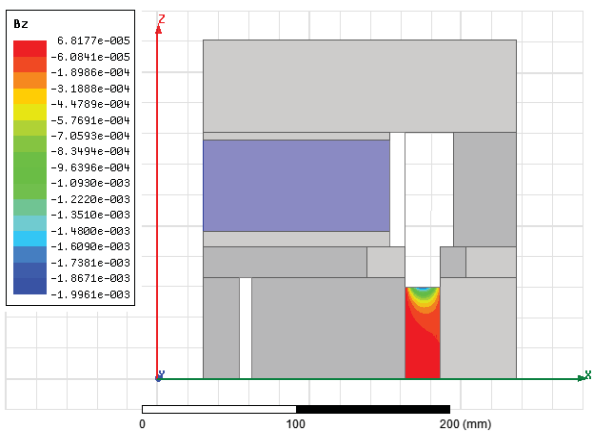

Fig. 8. The vertical component of the magnetic flux density in the required air gap when the misalign of the top and middle yoke is 0.1mm based on the design in figure 7 (a). The peak to peak value of $B_z$ in the required area is $2\times 10^{-3}$T.

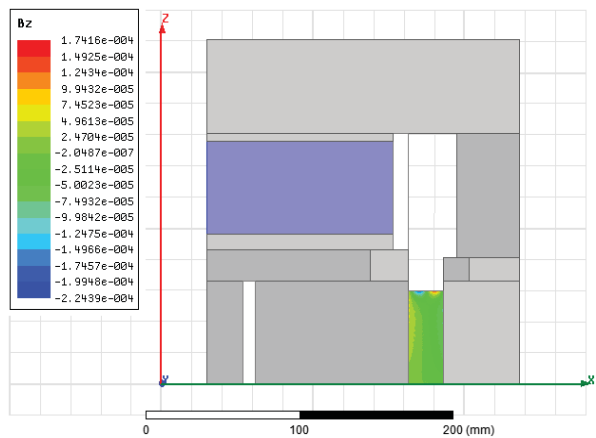

Fig. 9. The vertical component of the magnetic flux density in the required air gap when the misalign of the top and middle yoke is 0.1mm based on the design in figure 7 (b). The peak to peak value of $B_z$ in the required area is $4\times 10^{-4}$T.

IV. CONCLUSIONS

A designing model of the permanent magnet system for watt or joule balance is derived and how to adjust the uniformity of the magnetic flux density in the air gap is discussed. The analytical designing model directly expresses the physical principles behind the design and is a kind of guidance for the FEA software to find the optimal design. Based on this model, the permanent magnet system for NIM-2 is designed. A special design of opening shape makes the misalign of the top and middle yokes has little influence on the vertical component of the magnetic field.